\documentclass[reprint,amsmath,amssymb,aps,prl,superscriptaddress,email,floatfix]{revtex4-2}

\usepackage{header}
\begin{document}

\title{Resonant stroboscopic Rydberg dressing: electron-motion coupling and multi-body interactions}
\author{Chris Nill}
\affiliation{Institut f\"ur Theoretische Physik and Center for Integrated Quantum Science and Technology, Universit\"at Tübingen, Auf der Morgenstelle 14, 72076 T\"ubingen, Germany}
\affiliation{Institute for Applied Physics, University of Bonn, Wegelerstraße 8, 53115 Bonn, Germany}
\author{Sylvain de L\'es\'eleuc}
\affiliation{Institute for Molecular Science, National Institutes of Natural Sciences, 444-8585 Okazaki, Japan}
\affiliation{RIKEN Center for Quantum Computing (RQC), 351-0198 Wako, Japan}
\author{Christian Groß}
\affiliation{Physikalisches Institut and Center for Integrated Quantum Science and Technology, Universit\"{a}t T\"{u}bingen, Auf der Morgenstelle 14, 72076 T\"{u}bingen, Germany}
\author{Igor Lesanovsky}
\affiliation{Institut f\"ur Theoretische Physik and Center for Integrated Quantum Science and Technology, Universit\"at Tübingen, Auf der Morgenstelle 14, 72076 T\"ubingen, Germany}
\affiliation{School of Physics and Astronomy and Centre for the Mathematics and Theoretical Physics of Quantum Non-Equilibrium Systems, The University of Nottingham, Nottingham, NG7 2RD, United Kingdom}

\begin{abstract}
Rydberg dressing traditionally refers to a technique where interactions between cold atoms are imprinted through the far off-resonant continuous-wave excitation of high-lying Rydberg states. Dipolar interactions between these electronic states are then translated into effective interactions among ground state atoms. Motivated by recent experiments, we investigate two dressing protocols, in which Rydberg atoms are resonantly excited in a stroboscopic fashion. The first one is non-adiabatic, meaning Rydberg states are excited by fast pulses. In this case, mechanical forces among Rydberg atoms result in electron-motion coupling, which generates effective multi-body interactions. In the second, adiabatic protocol, Rydberg states are excited by smoothly varying laser pulses. We show that also in this protocol substantial multi-body interactions emerge.
\end{abstract}

\maketitle
\textit{Introduction ---}
Many current quantum simulation and computation platforms based on trapped ultracold atoms harness the properties of electronic Rydberg states \cite{Gallagher1994, Saffman2010, Bernien2017, Morgado2021,Browaeys2020, Lee2019, Levine2019, Magoni2023, Wilson2022}. These are highly excited states in which atoms are strongly polarizable and thus can interact over long distances. These interactions are typically orders of magnitude stronger than those between ground state atoms. Additionally, the short lifetime of Rydberg states is incompatible with the timescales of ultracold quantum dynamics. Rydberg dressing is a way to overcome this problem \cite{Jau2016, Zeiher2016}. In this process, the ground state of ultracold atoms is coupled to a Rydberg state via a far-detuned continuous-wave laser. This weak admixture of the Rydberg state enables sufficiently strong interactions and longer lifetimes, allowing for quantum simulations of novel phases, phase transitions and dynamics \cite{Fromholz2022, vanBijnen2015, Zeiher2017,Geissler2017,Khasseh2017,Guardado-Sanchez2018,Weckesser2024}. 

Recent developments have shown that stroboscopic dressing schemes may offer advantages over the conventional Rydberg dressing approach.
In stroboscopic dressing, Rydberg states are excited periodically using laser pulses, rather than through continuous off-resonant excitation.
This ensures that Rydberg states are populated only during discrete intervals, reducing overall interaction time and aligning the dynamics with the slower timescales of ultracold atoms.
Pulsed excitation protocols have also been studied in Refs.~\cite{Koyluoglu2024, Feldmeier2024} for realizing Floquet systems based on PXP models \cite{Fendley2004,Lesanovsky2012,Hudomal2022}, where the Rydberg blockade prevents the simultaneous excitation of nearby Rydberg atoms.

\begin{figure}
	\centering
	\includegraphics[width=\linewidth]{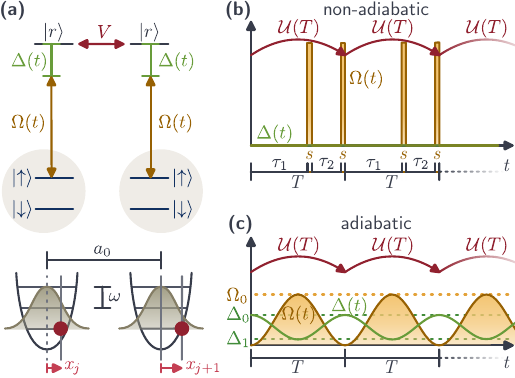}
	\caption{\textbf{Atomic levels and dressing protocols.}
	\textbf{(a)} 
	Two atoms trapped tweezers which are modeled as one-dimensional harmonic oscillators with trapping frequency $\omega$.
	Their equilibrium separation is $a_0$, and the displacement from their individual equilibrium positions is given by $x_j$ and $x_{j+1}$, respectively. 
	Each atom is modeled by two long-lived ground states $(\up, \down)$ and a Rydberg state \ry.
	Neighboring atoms in the state \ry interact with strength $V$.
	A globally applied laser couples \up and \ry with time-dependent Rabi frequency $\Omega(t)$ and detuning $\Delta(t)$.
	\textbf{(b)}
	The dressing protocols consist of periodic sequences of pulses of length $T$.
	In the non-adiabatic dressing protocol, two strong $\pi$~pulses of duration $s$ are applied, which break the Rydberg blockade, see Eq.~(\ref{eq:Rabi-frequence-non-adiabatic}).
	In between the two pulses, interacting Rydberg states are excited, which generates effective interactions among atoms in the state \up.
    \textbf{(c)} In the adiabatic dressing protocol, the Rabi frequency $\Omega(t)$ and the detuning $\Delta(t)$ are varied smoothly, see Eqs.~(\ref{eq:adiabatic-Rabi-frequency}) and  (\ref{eq:adiabatic-Detuning}). Rydberg states are excited during the pulses, but de-excited at the end. Dynamical phases accumulated during this process translate into effective interactions among atoms in the state \up.
	}
	\label{fig:cartoon}
\end{figure}

In this work, we study theoretically resonant stroboscopic Rydberg dressing in a cold atom tweezer array \cite{Browaeys2020}.
Specifically, we focus on two protocols, with the aim of creating effective spin Hamiltonians: the first involves fast pulses, which break the Rydberg blockade \cite{Lukin2001, Choi2007,Weidemuller2009,Pritchard2010,Gaetan2009,Barredo2014}.
In this case, the coupling between electronic excitation and the mechanical motion of atoms within their traps generates controllable spin-motion entanglement and multi-body interactions that can not be reduced to a sum of two-body terms \cite{Gambetta2020a,Bharti2024}.
The second protocol relies on slow adiabatic pulses. Here, the strength of the effectively generated multi-body interactions is controlled by the detuning of the Rydberg excitation laser and the spatial separation between the atoms.

\textit{General model ---}
We consider ultracold atoms of mass $m$ confined in optical tweezers as shown in Fig.~\ref{fig:cartoon}(a) in which they perform quantized oscillatory motion, characterized by the trap frequency $\omega$ and displacement coordinate $x_j$.
The equilibrium distance between two neighboring atoms is $a_0$.
The electronic degrees of freedom of each atom are modeled by two long-lived ground states (\down,\up), which form a fictitious spin, along with the Rydberg state \ry.
Rydberg atoms separated by the distance $x$ interact with the potential $V(x)$. Considering only interactions between nearest neighbors, the Hamiltonian of this system is given by
($\hbar=1$)
\begin{equation}
    H_0=\omega\sum_j a_j^\dagger a_j
    + V\sum_{j} n_j n_{j+1}
    + G\sum_{j} n_j n_{j+1} (x_j-x_{j+1}).
    \label{eq:H-Rydberg}
\end{equation}
Here $n_j=\dyad{r}_j$ is the projector on the Rydberg state of atom $j$, $V=V(a_0)$ is the interaction energy between nearest-neighbor Rydberg atoms and $G=\partial V(x)/\partial x|_{x=a_0}$ is the corresponding potential gradient \cite{Magoni2022}. The latter gives rise to a coupling between electronic and motional degrees of freedom, which are represented through raising and lowering operators: $x_j=x_0(a_j+a_j^\dagger)$ with the harmonic oscillator length $x_0=\sqrt{1/(2m\omega)}$.

The idea of Rydberg dressing is to generate effective interactions among atoms in the state \up, see Fig.~\ref{fig:cartoon}, by virtually or temporarily coupling this state to the Rydberg level \ry.
This coupling is achieved by a laser with time-dependent detuning (with respect to the \up-\ry transition) $\Delta(t)$ and Rabi frequency $\Omega(t)$.
This laser coupling is described by the Hamiltonian \cite{Cirac1992} 
\begin{equation}
    H_\mathrm{L}(t)=
    \Omega(t)\sum_j \left(\dyad{\uparrow}{r}_j e^{-i\kappa x_j}+\mathrm{h.c.}\right)
    +\Delta(t)\sum_j n_j,
    \label{eq:H-laser-general}
\end{equation}
where $\kappa$ is the projection of the laser wave vector onto the axis connecting neighboring atoms (we assume here that the atoms are position on a 1D chain in $x$-direction). In this work we consider stroboscopic dressing pulses, i.e. the laser parameters $\Delta(t)$ and $\Omega(t)$ are periodic functions with period $T$, as shown in Fig.~\ref{fig:cartoon}.
Our goal is to derive expressions for the time-evolution operator $\mathcal U(T)$ that propagates the system from the stroboscopic time $nT$ to $(n+1)T$. Rydberg states will only be excited during the dressing pulses, but (ideally) Rydberg excitations are present neither in the initial state nor in the states which the system assumes at the end of each pulse. Therefore, the time-evolution operator $U(T)$ only operates in the \up-\down manifold and is generated by an effective spin Hamiltonian, which we construct in the following for two different dressing protocols.

\textit{Non-adiabatic dressing protocol ---}
This protocol, which is shown in Fig.~\ref{fig:cartoon}(b), consists of two strong resonant $\pi$~pulses applied during a dressing cycle.
Each pulse has duration $s$, one is applied at time $\tau_1$ and the other at time $T-s$.
Mathematically, this sequence is modeled by choosing the parameters of the laser Hamiltonian (\ref{eq:H-laser-general}) as $\Delta(t)=0$ and 
\begin{equation}
    \Omega(t)=\frac{\pi}{2s}\left[\Pi_{\tau_1,\tau_1+s}(t)+\Pi_{T-s,T}(t)\right],
    \label{eq:Rabi-frequence-non-adiabatic}
\end{equation}
where $\Pi_{a,b}(t)=\Theta(t-a)-\Theta(t-b)$ and $\Theta$ is the Heaviside step function.
One dressing cycle can be decomposed as follows: The system starts at $t = 0$ in a spin state, $\ket{\psi_0}$, where no Rydberg excitations are present.
During the time interval $0\leq t\leq \tau_1$, the system evolves under Hamiltonian $H_0$, Eq.~(\ref{eq:H-Rydberg}), since the laser Hamiltonian $H_L$ is zero.
No Rydberg excitations are created, since $H_0$ conserves the total excitation number.
The first $\pi$~pulse excites all \up states to the Rydberg state \ry.
This assumes that the pulse is sufficiently strong in order to overcome the interaction among Rydberg atoms, i.e. it needs to break the Rydberg blockade, which is possible when $V\ll\pi/(2s)$.
Note, that this process may also lead to oscillatory excitations due to momentum transfer from the laser to the atoms.
In the following time interval of length $\tau_2$, nearest-neighbor atoms in the Rydberg state interact and further coupling between the electronic and motional degrees of freedom occurs.
The dressing cycle concludes with a second $\pi$~pulse, applied during the time interval $T-s\leq t \leq T$, that de-excites the Rydberg states.

The time evolution operator which propagates the system over a dressing cycle of duration $T$ can be expressed as a product of unitary operators. For a general pulse duration $s$, this is
\begin{equation}
    \mathcal U(T,s) =e^{-is (H_\mathrm L+H_0)}e^{-i H_0\tau_2}e^{-is (H_\mathrm L+H_0)}e^{-iH_0\tau_1}.
    \label{eq:non-adiabatic-exact}
\end{equation}
In the limit of infinitely fast $\pi$~pulses, i.e. $s\rightarrow 0$, it can be calculated analytically \cite{SM}, yielding
\begin{align}
    \mathcal{U}(T)&=\prod_j
    e^{-iT H_{j,\mathrm{eff}}}
    \mathcal{D}(\mathcal{J}_j)
    e^{-iT \omega a_j^\dagger a_j}.
    \label{eq:non-adiabatic-analytic-evolution}
\end{align}
In the following we analyze the terms of $\mathcal U(T)$ step by step. We begin with the last term, $e^{-i T \omega a_j^\dagger a_j}$, which is the simplest one and describes the free oscillatory motion of the atoms in their traps over the length $T$ of a dressing cycle.
The first term of $\mathcal{U}(T)$ is the evolution under the effective spin Hamiltonian $H_{j,\mathrm{eff}}$ which describes the time evolution of the spin at site $j$ during a dressing cycle as
\begin{align}
H_{j,\mathrm{eff}}=&\frac{\tau_2V}{T}\mathcal{P}_j \mathcal{P}_{j+1} 
     +\frac{\tau_2}{T}\frac{\mathcal{G}_j^2}{\omega}\left(\mathrm{sinc}(\omega\tau_2)-1\right)\nonumber\\
                    &+\frac{\eta^2\sin(\omega\tau_2)}{T}\mathcal{P}_j +\frac{\pi}{T}\mathcal{P}_j.
                    \label{eq:H-eff-general}
\end{align}
Here, the first term is the ``conventional'' dressing potential, i.e. the interaction $V$ among Rydberg atoms in Hamiltonian~(\ref{eq:H-eff-general}), is mapped onto atoms in the \up-state but with weaker effective strength $V_\mathrm{eff}=\tau_2 V /T$.
The factor of $\tau_2/T$ is the proportion of the time during which the system is in the Rydberg state during a dressing cycle.
Further terms emerge in the effective spin Hamiltonian, with the next one being an interaction that is mediated by the oscillation of the atoms in their tweezer traps.
It depends on the gradient $G$ and the oscillator length $x_0$ through the operator
\begin{align}
   \mathcal{G}_j=\bigg\{
   \begin{array}{lr}
        Gx_0\mathcal{P}_j\mathcal{P}_{j+1}, & \text{for } N=2,\\
        Gx_0\mathcal{P}_j\left(\mathcal{P}_{j+1} -\mathcal{P}_{j-1} \right), & \text{for } N>2.
        \end{array}
\end{align}
For two atoms, this contribution, which is generated by the coupling between electronic and motional degrees of freedom, merely modifies the interaction strength between atoms in the \up-state.
For more than two atoms, however, it results in a multi-body interaction \cite{Gambetta2020,Nill2022}.
This can be understood as follows: when an atom $j$ is surrounded by two or no atoms in the \up-state, the net mechanical force acting on it when excited to the \up-state is zero. 
Hence $\mathcal{G}_j=0$.
For other configurations, $\mathcal{G}_j$ assumes a non-zero value.
Distinguishing these cases necessitates an operator which acts on site $j$ as well as on the two neighboring sites, hence generating a three-body interaction.
The term $\eta^2\sin(\omega\tau_2)\mathcal{P}_j/T$ in the effective Hamiltonian accounts for the momentum transfer during laser excitation. It is quantified by the Lamb-Dicke parameter $\eta=\kappa x_0$ \cite{Ryabtsev2011,Eschner2003, Stenholm1986}.
The last term, $\frac{\pi}{T} \mathcal{P}_j$ arises due to the application of two $\pi$~pulses during the dressing cycle, which leads to a phase shift of $\pi$ for all atoms in the state \up.

Let us now return to the middle term of the unitary evolution operator (\ref{eq:non-adiabatic-analytic-evolution}). It depends on the displacement operator $\mathcal D(\mathcal{J}_j) = \exp{\mathcal J_j a_j^\dagger - \mathcal J_j^\dagger a_j}$ with argument
$\mathcal{J}_j=\left(\frac{\mathcal{G}_j}{\omega}+i\eta \mathcal{P}_j\right) \left(e^{-i\omega\tau_2}-1\right)$.
It displaces atom $j$ in position (momentum) space by the real (imaginary) part of $\mathcal J_j$, which is an operator that acts on the spin degrees of freedom \cite{Cahill1969}.
This term therefore results in spin-motion coupling.
The amount of the displacement further depends on the oscillation phase $\omega\tau_2$, which can be utilized to control the spin-motion coupling.
If the oscillation phase is zero or a multiple of $2\pi$, $\mathcal{J}_{j}$ vanishes and spin and motional degrees of freedom decouple.

\begin{figure}[t]
    \centering
    \includegraphics[width=\linewidth]{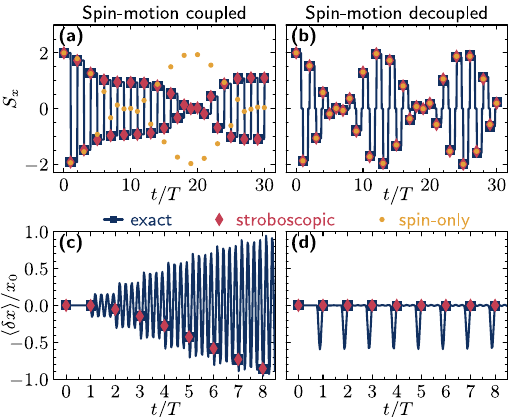}
    \caption{\textbf{Non-adiabatic dressing protocol.}
    \textbf{(a)}~Evolution of the expectation value of the total spin component $S_x$ for two atoms (see text). Choosing $\omega \tau_2 = 0.01\pi$, the oscillatory motion is coupled to the spin dynamics. This leads to entanglement between the two degrees of freedom, causing spin decoherence (see Supplemental Material \cite{SM}). Numerically exact results calculated by $\mathcal U(T,s)$, Eq.~(\ref{eq:non-adiabatic-exact}), are shown as blue squares and the stroboscopic evolution under the unitary $\mathcal{U}(T)$, Eq.~(\ref{eq:non-adiabatic-analytic-evolution}), is shown as red diamonds. For comparison, we show the evolution of $S_x$ in the absence of this spin-motion coupling calculated by Eq.~(\ref{eq:decoupled-spin-evolution}) with yellow dots. The initial state $\ket{\psi_0}$ is defined in the main text.
    \textbf{(b)}~When choosing $\omega \tau_2 = 2\pi$ spin and motion decouple.
    \textbf{(c)}~Dynamics of the average interatomic distance $\ev{\delta x}$ (see text). For $\omega \tau_2 = 0.01\pi$, spin-motion coupling results in oscillations of the interatomic distance after a dressing cycle.
    \textbf{(d)}~When choosing $\omega \tau_2 = 2\pi$, the motional degrees of freedom return to their initial state after each cycle.
    In all simulations the harmonic oscillator basis is truncated to contain a maximum of $5$ excitations, the parameters values are $T=8.1\pi/\omega$, $V = 10.1\omega$, $\eta = 0.6$ and $s = \pi/(2\cross 10^4\omega)$.
    For panels (a,c) we use $G x_0 = 5\omega$ and for panels (b,d) we set $Gx_0 = 0.3\omega$.
    Note that in experiment, the values for $V$ and $G$ can in fact be tuned independently using microwave dressing of Rydberg states \cite{Petrosyan2014,Sevincli2014,Scholl2022,Kastner2024}.
    }
    \label{fig:non-adiabatic-evolution}
\end{figure}

For a dressing cycle with such decoupled spin and motion, the time evolution operator separates into an operator acting on the spins,
\begin{align}
    \mathcal{U}_\mathrm{spin}(T)
    &=\prod_j
    e^{-iT H_{j,\mathrm{eff}}},
    \label{eq:decoupled-spin-evolution}
\end{align}
and one acting on the harmonic oscillatory motion of the atoms, $\mathcal{U}_\mathrm{osc}=\prod_j e^{-iT\omega a_j^\dagger a_j}$.
However, multi-body interactions still remain in the effective spin Hamiltonian due to its dependence on $\mathcal G_j^2$.

To illustrate the effect of the spin-motion coupling, we consider in  Fig.~\ref{fig:non-adiabatic-evolution} a system of two atoms which are initialized in the motional ground state $\ket{0}_j$.
As initial state we choose
$\ket{\psi_0}=\bigotimes_{j=1}^{N=2} \left[\frac{1}{\sqrt 2}\left(\up_j+\down_j\right)\otimes\ket{0}_j\right]$, which is not an eigenstate of the effective spin Hamiltonian and thus leads to non-trivial dynamics.
In the figure, we study the dynamics of two observables of the system:
the expectation value of the $x$-component of the total spin, $S_x=\sum_{j=1}^{N=2}\ev{\dyad{\uparrow}{\downarrow}_j+\dyad{\downarrow}{\uparrow}_j}$, and the expectation value of the distance $\ev{\delta x}/x_0=\ev{x_2-x_1}/x_0$ between the atoms.
The time evolution of the observables is calculated numerically exactly through $\mathcal U(T,s)$ for finite $\pi$~pulse length $s$, Eq.~(\ref{eq:non-adiabatic-exact}), and compared to the time evolution operator $\mathcal U(T)$ in the limit $s\rightarrow 0$, Eq.~(\ref{eq:non-adiabatic-analytic-evolution}).
For $\omega\tau_2=0.1\pi$ spins and motion couple, whereas both degrees of freedom perform a separable evolution when $\omega\tau_2=2\pi$.
This is shown in Fig.~\ref{fig:non-adiabatic-evolution}(a,b), where we display the behavior of spin observable $S_x$ as a function of time for both cases.
Spin-motion coupling also affects the evolution of the distance $\ev{\delta x}$ between the atoms: With spin-motion coupling, the distance varies strongly at each stroboscopic time step $nT$, while it remains constant in the decoupled case, see Fig.~\ref{fig:non-adiabatic-evolution}(c),(d).
To approximately decouple spin and motion, one can choose very short pulse durations, such that $\omega\tau_2\ll1$.
Dressing with such ultra-strong laser pulses has been experimentally demonstrated in Ref.~\cite{Bharti2024}. We discuss this special case in the Supplemental Material \cite{SM}. There, we also illustrate how a spin-motion echo sequence can decouple spin and motional dynamics.

\textit{Adiabatic dressing protocol ---}
This dressing protocol employs slowly changing laser parameters rather than fast pulses. It is sketched in Fig. \ref{fig:cartoon}(c).
The time evolution during a dressing cycle of length $T$ is assumed to be adiabatic, in the sense that the spin state at its beginning and its end is the same. Moreover, the state shall be smoothly connected to the instantaneous eigenstates during the adiabatic pulse, which contain components of Rydberg states. This necessity to be slow, i.e. to avoid non-adiabatic transitions, is in conflict with the finite lifetimes of Rydberg states. Hence, an appropriate parameter regime needs to be identified that ensures pulses which are much shorter than the Rydberg atom's lifetime. This is indeed possible, as shown in the Supplemental Material \cite{SM}.
For simplicity, and to make the calculations tractable, we neglect in the following the coupling between the electronic degrees of freedom and the oscillatory motion, setting the potential gradient $G$ and the Lamb-Dicke parameter $\eta$ to zero. 

\begin{figure}
    \centering
    \includegraphics{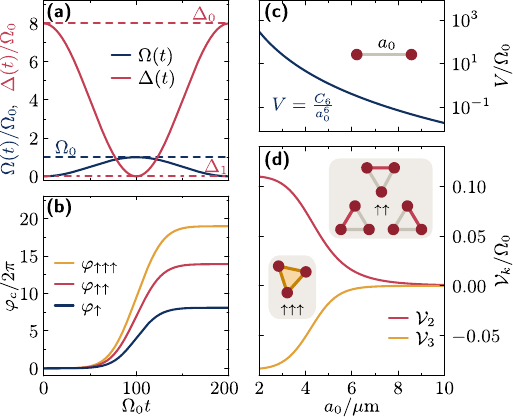}
    \caption{\textbf{Adiabatic dressing protocol.}
    \textbf{(a)}~Time-dependent Rabi frequency $\Omega(t)$ and detuning $\Delta(t)$ for a dressing cycle with duration $T=200/\Omega_0$, where $\Delta_0=8\Omega_0$ and $\Delta_1=0$.
    \textbf{(b)}~Accumulated dynamical phase $\varphi_{\mathcal C}$ in a system of three atoms placed on an equilateral triangle. Here, the initial configurations ${\mathcal C}$ need to be distinguished only by their number of atoms in the \up-state. The interaction strength is $V=4\Omega_0$.
    \textbf{(c)}~Bare van der Waals interaction between Rydberg atoms as a function of the interatomic distance $a_0$. We use the dispersion coefficient $C_6=\SI{119}{\giga\hertz\,\micro\meter^6}$ of state $61\mathrm{S}_{1/2}$ of potassium with Rabi frequency $\Omega_0=2\pi\cross\SI{1}{\mega\hertz}$.
    \textbf{(d)}~Effective two- and three-body interaction strength, $\mathcal V_2$ and $\mathcal{V}_3$, for varying interatomic distance $a_0$. The calculation is based on the bare van der Waals interaction shown in panel (c).
    The closer interatomic distance, the higher is the relative contribution of the three-body interaction $\mathcal{V}_3$.
    }
    \label{fig:adiabatic_sample}
\end{figure}

In the adiabatic dressing protocol, the Rabi frequency and detuning are smoothly varied according to Fig.~\ref{fig:adiabatic_sample}(a)~as
\begin{align}
    \Omega(t)&=\Omega_0\sin^2\left(\pi t/T\right)
    \label{eq:adiabatic-Rabi-frequency}\\
    \Delta(t)&=\Delta_0 - (\Delta_0-\Delta_1)\sin^2(\pi t / T).
    \label{eq:adiabatic-Detuning}
\end{align}
The dynamical phase that is accumulated during this pulse differs for each spin configuration and yields an effective Hamiltonian in the spin space which in general contains multi-body interactions. 
As an example, we consider a system with $N=3$ atoms positioned in an equilateral triangle whose sides have the length $a_0$.
The effective spin Hamiltonian can generally be written in terms of the projectors $\mathcal P_j=\dyad{\uparrow}_j$:
\begin{equation}
    H_\mathrm{eff}=
          \mathcal V_\mathrm{3} \mathcal P_1 \mathcal P_2 \mathcal P_3
        + \mathcal V_2\sum_{i\neq j} \mathcal P_i \mathcal P_j
        + \mathcal V_1\sum_i \mathcal P_i
        + \mathcal V_0,
        \label{eq:ad_H_eff}
\end{equation}
where $\mathcal V_{k}$ describes the effective $k$-body interaction strength.
The evolution under this effective spin Hamiltonian over a time $T$ then generates the dynamical phase $\varphi_\mathcal C$ of the spin configuration $\mathcal C\in\left\{\downarrow\downarrow\downarrow,\downarrow\downarrow\uparrow,\dots,\uparrow\uparrow\uparrow\right\}$:
\begin{equation}
    e^{-iH_\mathrm{eff}T}=\sum_\mathcal C e^{-i\varphi_\mathcal C}\dyad{\mathcal C}.
\end{equation}
Comparing coefficients, we can now link the $k$-body interaction strength to the dynamical phases. Due to the symmetry of the system, the phases $\varphi_\mathcal C$ depend only on $\up$-spins that are contained in configuration $\mathcal{C}$. We thus use a shorthand notation $\varphi_\alpha$, where the subscript $\alpha$ refers to the number of atoms in the \up-state, i.e. we have $\varphi_{\downarrow\downarrow\downarrow}=\varphi_0$, $\varphi_{\uparrow\downarrow\downarrow}=\varphi_{\downarrow\uparrow\downarrow}=\varphi_{\downarrow\downarrow\uparrow}=\varphi_\uparrow$, etc. With this, we obtain
\begin{align}
    \mathcal V_0 T  &= \varphi_{0},
    &\mathcal V_1 T  &= \varphi_{\uparrow}-\varphi_{0},\label{eq:effective-Vs}\\
    \mathcal V_2 T  &= \varphi_{\uparrow\uparrow}-2\varphi_{\uparrow}+\varphi_{0},
    &\mathcal V_3 T  &= \varphi_{\uparrow\uparrow\uparrow}-3\varphi_{\uparrow\uparrow}+3\varphi_{\uparrow}-\varphi_{0}.\nonumber
\end{align}
In the following, we construct the effective spin Hamiltonian~(\ref{eq:ad_H_eff}) for an exemplary case and show that indeed rather generically three-body interactions emerge.

Using numerical integration \cite{Johansson2013}, we solve for each initial state from $\mathcal{C}$ the time-dependent Schrödinger equation with the Hamiltonian $H(t)=H_0+H_\mathrm L(t)$ [Eqs.~(\ref{eq:H-Rydberg}), (\ref{eq:H-laser-general})] and the adiabatic pulse~(\ref{eq:adiabatic-Rabi-frequency}),(\ref{eq:adiabatic-Detuning}).
We then extract the adiabatically accumulated dynamical phase $\varphi_\mathcal{C}$ as a function of the laser parameters and the interaction strength $V$. An example is shown in Fig.~\ref{fig:adiabatic_sample}(b). From these phases, we can then calculate the interaction coefficients $\mathcal{V}_k$ using the above-described procedure. We assume here that the bare interaction between atoms in Rydberg states is given by a van der Waals interaction of the form $V=C_6/a_0^6$, see Fig.~\ref{fig:adiabatic_sample}(c), where $C_6$ is the dispersion coefficient \cite{Beguin2013}. Based on this, we calculate the effective $k$-body interaction strength as a function of the interatomic distance $a_0$, which is shown in panel (d). This results in two- and three-body potentials, which have a characteristic flat-top form \cite{Balewski2014}. The larger the interparticle separation, i.e. the smaller the bare interaction $V$, the more dominant becomes the two-body interaction. A detailed analysis of the ratio $\mathcal{V}_3/\mathcal V_2$ is outlined in the Supplemental Material~\cite{SM}.

\textit{Conclusions and outlook --- }
We discussed the theory underlying two stroboscopic Rydberg dressing protocols for generating effective spin Hamiltonians. The non-adiabatic protocol relies on the application of fast resonant laser pulses which break the Rydberg blockade. This results in state-dependent mechanical forces causing a coupling between motional and electronic degrees of freedom. As a consequence effective multi-body spin interactions are generated, but also spin decoherence emerges which can be mitigated by an appropriate timing of the laser pulses. The adiabatic protocol, on the other hand, is implemented by smoothly varying laser pulses. During the adiabatic evolution, the system probes interacting configurations of Rydberg atoms, which generate spin-dependent dynamical phases resulting in effective $k$-body spin interactions. For a system of three atoms, we analyzed two-body and three-body interactions and their dependence on the interparticle distance.

There are a number of avenues for interesting future work. The considerations of this paper are based on simple 1D and 2D systems, and the effect of different lattice geometries and multidimensional oscillatory excitations certainly merits further exploration. Additionally, quantum optimal control techniques may allow engineering stroboscopic dressing protocols that reduce spin-motion coupling or enhance adiabaticity while minimizing spin decoherence. In particular, it would be interesting to explore whether two-body interactions could be completely removed, leading to many-body Hamiltonians that feature solely multi-body interactions.

\FloatBarrier
\acknowledgments
\textit{Acknowledgments ---}
We thank Wilson Martins, Albert Cabot and Joseph Wilkinson for fruitful discussions.
We acknowledge funding from the Deutsche Forschungsgemeinschaft within SPP 1929 GiRyd (Grant No. 428276754: LE3522/1 and GR4741/5), project GR4741/4 and the research units FOR5413 (Grant No. 465199066) and FOR5522 (Grant No. 499180199). We also acknowledge funding from the Horizon Europe programme HORIZON-CL4-2022-QUANTUM-02-SGA via the project 101113690 (PASQuanS2.1), the Baden-Württemberg Stiftung through Project No.~BWST\_ISF2019-23, the Alfried Krupp von Bohlen and Halbach foundation and the state of Baden-Württemberg through bwHPC grant no INST 40/575-1 FUGG (JUSTUS 2 cluster). SdL acknowledges funding from the MEXT Quantum Leap Flagship Program (MEXT Q-LEAP) JPMXS0118069021 and JST Moonshot R\&D Program Grant Number JPMJMS2269. CG and SdL acknowledge funding from the NINS-DAAD exchange program (Quantum Simulation/Computation: From Rydberg Dressing to Ultrafast Interaction).

\bibliography{MeineBibliothek}


\clearpage

\setcounter{equation}{0}
\setcounter{figure}{0}
\setcounter{table}{0}
\setcounter{secnumdepth}{2}
\makeatletter
\renewcommand{\theequation}{S\arabic{equation}}
\renewcommand{\thefigure}{S\arabic{figure}}

\makeatletter
\renewcommand{\theequation}{S\arabic{equation}}
\renewcommand{\thefigure}{S\arabic{figure}}

\onecolumngrid
\setcounter{page}{1}

\begin{center}
{\Large SUPPLEMENTAL MATERIAL}
\end{center}
\begin{center}
\vspace{0.8cm}
{\Large Resonant stroboscopic Rydberg dressing: electron-motion coupling and multi-body interactions}
\end{center}
\begin{center}
Chris Nill$^{1,2}$, Sylvain de L\'es\'eleuc$^{3,4}$, Christian Groß$^{5}$, and Igor Lesanovsky$^{1,6}$
\end{center}
\begin{center}
    $^1${\it Institut f\"ur Theoretische Physik and Center for Integrated Quantum Science and Technology, Universit\"at T\"ubingen,}
    {\it Auf der Morgenstelle 14, 72076 T\"ubingen, Germany}\\
    $^2${\it Institute for Applied Physics, University of Bonn, Wegelerstraße 8, 53115 Bonn, Germany}\\
    $^3${\it Institute for Molecular Science, National Institutes of Natural Sciences, 444-8585 Okazaki, Japan}\\
    $^4${\it RIKEN Center for Quantum Computing (RQC), 351-0198 Wako, Japan}\\
    $^5${\it Physikalisches Institut and Center for Integrated Quantum Science and Technology, Universit\"{a}t T\"{u}bingen, Auf der Morgenstelle 14, 72076 T\"{u}bingen, Germany}\\
    $^6${\it School of Physics and Astronomy and Centre for the Mathematics and Theoretical Physics of Quantum Non-Equilibrium Systems, The University of Nottingham, Nottingham, NG7 2RD, United Kingdom}
\end{center}
\tableofcontents

\section{Analysis of the non-adiabatic dressing protocol}
\subsection{Derivation of the time evolution operator \texorpdfstring{$\mathcal U(T)$}{U(T)} at stroboscopic times}
The non-adiabatic dressing sequence, see Fig.~1(b) in the main text, is decomposed into four phases. During the first one of duration $\tau_1$, the system is described by the free evolution under the Rydberg Hamiltonian ($\hbar=1$)
\begin{equation}
    H_0=\omega\sum_j a_j^\dagger a_j + V\sum_j n_jn_{j+1}+G\sum_j n_jn_{j+1}(x_j-x_{j+1}),
    \label{eq:H0}
\end{equation}
where $\omega$ is the trapping frequency of the atoms with harmonic oscillator operators $a_j$ and $a_j^\dagger$. The Rydberg interaction strength is parametrized by $V$, and the gradient of the potential by $G$. The projector onto an atom in Rydberg state $\ket r $ is given by $n_j=\dyad{r}_j$.

During the second phase of duration $s$, the laser Hamiltonian with Rabi frequency $\Omega$ and projected wave vector $\kappa$ (the axis connecting the two atoms is the chosen as $x$-axis) is acing. This leads to the additional Hamiltonian
\begin{equation}
    H_L=\sum_j\Omega\left(\dyad{\uparrow}{r}_j e^{-i\kappa x_j}+\dyad{r}{\uparrow}_j e^{i\kappa x_j}\right),
    \label{eq:H-realistic-laser}
\end{equation}
where the position operator of atom $j$ with atomic mass is
\begin{equation}
    x_j=x_0(a_j+a_j^\dagger)=\sqrt{\frac{1}{2m\omega}}(a_j+a_j^\dagger).
\end{equation}
The Lamb-Dicke parameter is then given as $\eta=\kappa x_0$.
During the third phase of the dressing sequence, with duration $\tau_2$, only $H_0$ is acting (the laser is switched off), while during he fourth phase is the same as the second.
Consequently, the time-evolution operator of a sequence of length $T=\tau_1+\tau_2+2s$ is given by
\begin{equation}
    \mathcal{U}_{\tau_1+\tau_2+2s}=\underbrace{e^{-is (H_L+H_0)}}_{\mathcal{U}_\pi}e^{-i H_0\tau_2}e^{-is (H_L+H_0)}e^{-i H_0\tau_1}.
    \label{eq:time-evolution-first}
\end{equation}
First, we evaluate now
\begin{align}
    \mathcal{U}_\pi&=e^{-is (H_L+H_0)}\\
    &=\exp{-is \Omega\sum_j \left(\dyad{\uparrow}{r}_j e^{-i\kappa x_j}+\dyad{r}{\uparrow}_j e^{i\kappa x_j}+\frac{\omega}{\Omega} a_j^\dagger a_j
    +\frac{V}{\Omega} n_jn_{j+1}
    + \frac{G_j}{\Omega}n_jn_{j+1}(x_j-x_{j+1})\right)}.
\end{align}
We consider a strong laser driving which breaks the Rydberg blockade and is by magnitudes the largest energy scale with $V/\Omega\ll1,\, \omega/\Omega\ll 1,\, G_j/\Omega\ll 1$, thus
\begin{align}
    \mathcal{U}_\pi
    &= e^{-isH_L}
    =\exp{-is \Omega\sum_j \left(\dyad{\uparrow}{r}_j e^{-i\kappa x_j}+\dyad{r}{\uparrow}_j e^{i\kappa x_j}\right)}.
    \label{eq:pi-pulse-approx}
\end{align}
In addition, $\Omega s=\pi/2$ is chosen such that the laser performs $\pi$~pulses, which reduce effectively to instantaneous pulses
\begin{align}
    \mathcal{U}_\pi
    &=\prod_j\exp{-i\Omega s\left(\dyad{\uparrow}{r}_j e^{-i\kappa x_j}+\dyad{r}{\uparrow}_j e^{i\kappa x_j}\right)+0\cdot\dyad{\downarrow}_j}\\
    &=\prod_j\left[\dyad{\downarrow}_j-i\left(\dyad{\uparrow}{r}_j e^{-i\kappa x_j}+\dyad{r}{\uparrow}_j e^{i\kappa x_j}\right)\right].
\end{align}
The time-evolution operator from Eq.~(\ref{eq:time-evolution-first}) for a sequence reads then
\begin{equation}
    \mathcal{U}_{\tau_1+\tau_2+2s}=\underbrace{\mathcal{U}_\pi e^{-i H_0\tau_2}\mathcal{U}_\pi}_\mathcal{U_R} e^{-i H_0\tau_1}.
\end{equation}
The expression $\mathcal{U_R}$ can be evaluated as
\begin{align}
    \mathcal{U_R}
    &=\mathcal{U}_\pi e^{-i H_0\tau_2}\mathcal{U}_\pi
    =\underbrace{\mathcal{U}_\pi\mathcal{U}_\pi}_{\prod_j(\dyad{\downarrow}_j-\mathcal{P}_j -n_j)}\mathcal{U}_\pi^\dagger e^{-i H_0\tau_2}\mathcal{U}_\pi\\
    &=\prod_j\left(\dyad{\downarrow}_j-\mathcal{P}_j-n_j\right)
        \exp{-i\tau_2\left(\omega\sum_j \underbrace{\mathcal{U}_\pi^\dagger a_j^\dagger a_j\mathcal{U}_\pi}_A
        + V\sum_j \underbrace{\mathcal{U}_\pi^\dagger n_jn_{j+1}\mathcal{U}_\pi}_{\mathcal{P}_j\mathcal{P}_{j+1}}
        +\sum_j \underbrace{G_j\mathcal{U}_\pi^\dagger n_jn_{j+1}(x_j-x_{j+1})\mathcal{U}_\pi}_{\mathcal{P}_j\mathcal{P}_{j+1}(x_j-x_{j+1})},
    \right)},
\end{align}
where we define the projector onto atom $j$ in state $\ket \uparrow$ as $\mathcal{P}_j=\dyad{\uparrow}_j$.
In addition,
\begin{equation}
    A=\omega \left(a_j^\dagger a_j +i\eta (n_j-\mathcal{P}_j)(a_j-a_j^\dagger)+\eta^2(n_j+\mathcal{P}_j)\right).
\end{equation}
The time-evolution operator for a dressing sequence then is
\begin{align}
    \mathcal{U}_{\tau_1+\tau_2+2s}
    =&\left[\prod_j\left(\dyad{\downarrow}_j-\mathcal{P}_j-n_j\right)\right]\nonumber\\
        &\cdot\exp{-i\tau_2\left(\omega\sum_j \left(a_j^\dagger a_j +i\eta (n_j-\mathcal{P}_j)(a_j-a_j^\dagger)+\eta^2(n_j+\mathcal{P}_j)\right)
        +\sum_j G_j\mathcal{P}_j\mathcal{P}_{j+1}(x_j-x_{j+1})
        \right)}\nonumber\\
    &\cdot\exp{-i\tau_1\left(\omega\sum_j a_j^\dagger a_j
        +\sum_j G_j n_jn_{j+1}(x_j-x_{j+1})
        \right)}\nonumber\\
    &\cdot\exp{-i\tau_2 V\sum_j \mathcal{P}_j \mathcal{P}_{j+1}}
        \cdot\exp{-i\tau_1 V\sum_j n_j n_{j+1}}
        .
\end{align}

We consider an initial state $\ket{\psi_0}$ in which no Rydberg excitations are present.
Thus, multiplying $\prod_j (\mathds 1-n_j)$ onto the time evolution does not affect the initial state.
Since $\mathcal{U}_{\tau_1+\tau_2+2s}$ commutes with $n_j$, we can write
\begin{align}
    \mathcal{U}_{\tau_1+\tau_2+2s}
    =&\,\,\mathcal{U}_{\tau_1+\tau_2+2s}\cdot \prod_j (\mathds{1}-n_j)\\
    =&\prod_j\left(\dyad{\downarrow}_j-\mathcal{P}_j\right)
        \exp{-i\tau_2\left(\omega\sum_j \left(a_j^\dagger a_j
        -i\eta \mathcal{P}_j(a_j-a_j^\dagger)+\eta^2\mathcal{P}_j\right)
        +\sum_j G_j\mathcal{P}_j\mathcal{P}_{j+1}(x_j-x_{j+1})
        \right)}\nonumber\\
        &\cdot\exp{-i\tau_1\omega\sum_j a_j^\dagger a_j}
        \cdot\exp{-i\tau_2 V\sum_j \mathcal{P}_j \mathcal{P}_{j+1}}\\
    =&\prod_j\left(\dyad{\downarrow}_j-\mathcal{P}_j\right)\nonumber\\
    &\cdot\textcolor{AccentsBlue5}{\underbrace{\exp{-i\tau_2\left(\omega\sum_j \left(a_j^\dagger a_j
    -i\eta \mathcal{P}_j(a_j-a_j^\dagger)+\eta^2\mathcal{P}_j\right)
    +\sum_j G_j\mathcal{P}_j\mathcal{P}_{j+1}(x_j-x_{j+1})
    \right)}\cdot\exp{i\tau_2\omega\sum_j a_j^\dagger a_j}}_\text{spin-motion coupling dynamics}}\nonumber\\
    &\textcolor{AccentsYellow4}{\cdot\underbrace{\exp{-i(\tau_1+\tau_2)\omega\sum_j a_j^\dagger a_j}}_\text{free-oscillator dynamics}}
    \textcolor{AccentsRed4}{\underbrace{\cdot\exp{-i(\tau_1+\tau_2) \frac{\tau_2 V}{\tau_1+\tau_2}\sum_j \mathcal{P}_j \mathcal{P}_{j+1}}}_\text{spin-spin dynamics}}
    \label{eq:time-evolution-2}
    .
\end{align}
\mybox{
    \textbf{Gradient operator $\mathcal{G}_j$}\\
    Introducing the gradient operator enables us to commute the summands with
    \begin{align}
         \sum_j G_j \mathcal{P}_j \mathcal{P}_{j+1}(x_j-x_{j+1})
        =&\sum_j G_j \mathcal{P}_j \mathcal{P}_{j+1} x_j-\sum_j G_j \mathcal{P}_j \mathcal{P}_{j+1} x_{j+1}\nonumber\\
        =&\sum_j G_j \mathcal{P}_j \mathcal{P}_{j+1} x_j-\sum_j G_{j-1} \mathcal P_{j-1} \mathcal{P}_j x_j\nonumber\\
        =&\sum_j \left(G_j\mathcal{P}_{j+1}-G_{j-1}\mathcal P_{j-1}\right)\mathcal{P}_jx_j\nonumber\\
        =&\sum_j \left(G_j\mathcal{P}_{j+1}-G_{j-1}\mathcal P_{j-1}\right) \mathcal{P}_jx_0(a_j+a_j^\dagger)\nonumber\\
        \eqqcolon&\sum_j\mathcal{G}_j (a_j+a_j^\dagger).
        \label{eq:surrounding-G-operator}
    \end{align}
    Note, that $\comm{\mathcal{G}_j}{\mathcal{G}_l}=0\,\forall j,l$ and $\mathcal{G}_j^2\neq\mathcal{G}_j$.
}{AccentsRed3}
The spin-motion term (blue) can be further simplified using the gradient operator $\mathcal{G}_j$, see Eq.~(\ref{eq:surrounding-G-operator}), as
\begingroup
\allowdisplaybreaks
\begin{align}
    &\exp{-i\tau_2\left(
            \omega\sum_j \left(a_j^\dagger a_j -i\eta \mathcal{P}_j(a_j-a_j^\dagger)+\eta^2\mathcal{P}_j\right)
            +\sum_j G_j\mathcal{P}_j\mathcal{P}_{j+1}(x_j-x_{j+1})
        \right)}
        \cdot\exp{i\tau_2\omega\sum_j a_j^\dagger a_j}\\
    \intertext{and since all terms commute now in $j$}\nonumber\\
    &=\prod_j\exp{-i\tau_2\left(
            \omega a_j^\dagger a_j-i\eta\omega \mathcal{P}_j(a_j-a_j^\dagger)+\eta^2\omega \mathcal{P}_j
            +\mathcal{G}_j(a_j+a_j^\dagger)
        \right)}
        \cdot\exp\left\{i \tau_2\omega a_j^\dagger a_j\right\}.\\
    \intertext{We complete the square with}\nonumber\\
    &=\prod_j\exp{-i\tau_2\omega\left[
            \Big(a_j^\dagger
                \underbrace{-i\eta \mathcal{P}_j+\frac{\mathcal{G}_j}{\omega}}_{\mathcal{J}_j^\dagger}\Big)
            \Big(a_j+\underbrace{i\eta \mathcal{P}_j+\frac{\mathcal{G}_j}{\omega}}_{\mathcal{J}_j}\Big)
            +\eta^2\mathcal{P}_j-\mathcal{J}_j^\dagger\mathcal{J}_j
        \right]}
        \cdot\exp\left\{i \tau_2\omega a_j^\dagger a_j\right\}\\
    &=\prod_j\exp{-i\tau_2\omega\left[
            (a_j+\mathcal{J}_j)^\dagger
            (a_j+\mathcal{J}_j)
            +\eta^2\mathcal{P}_j-\mathcal{J}_j^\dagger\mathcal{J}_j
        \right]}
        \cdot\exp\left\{i \tau_2\omega a_j^\dagger a_j\right\}\\
    \intertext{and introduce the displacement operator with $a+\alpha=\mathcal{D}^\dagger(\alpha)a\mathcal{D}(\alpha)$, thus}\nonumber\\
    &=\prod_j\exp{-i\tau_2\omega\left[
            \left(\mathcal{D}^\dagger(\mathcal{J}_j)a_j\mathcal{D}(\mathcal{J}_j)\right)^\dagger
            \mathcal{D}^\dagger(\mathcal{J}_j)a_j\mathcal{D}(\mathcal{J}_j)
            +\eta^2\mathcal{P}_j-\mathcal{J}_j^\dagger\mathcal{J}_j
        \right]}
        \cdot\exp\left\{i \tau_2\omega a_j^\dagger a_j\right\}\\
    &=\prod_j\exp{-i\tau_2\omega\left[
            \mathcal{D}^\dagger(\mathcal{J}_j)a^\dagger\underbrace{\mathcal{D}(\mathcal{J}_j)
            \mathcal{D}^\dagger(\mathcal{J}_j)}_{\mathds{1}}a\mathcal{D}(\mathcal{J}_j)
            +\eta^2\mathcal{P}_j-\mathcal{J}_j^\dagger\mathcal{J}_j
        \right]}
        \cdot\exp\left\{i \tau_2\omega a_j^\dagger a_j\right\}.\\
    \intertext{Note that all terms in $\exp$ commute now, thus}\nonumber\\
    &=\prod_j\exp{-i\tau_2\omega\left[
            \mathcal{D}^\dagger(\mathcal{J}_j)a_j^\dagger a_j\mathcal{D}(\mathcal{J}_j)
        \right]}
        \cdot\exp\left\{i \tau_2\omega a_j^\dagger a_j\right\}
        \cdot\exp\left\{-i\tau_2\omega \left(\eta^2\mathcal{P}_j-\mathcal{J}_j^\dagger\mathcal{J}_j\right)\right\}.\\
    \intertext{Since $\mathcal{D}(\mathcal{J}_j)$ is a unitary operator and thus invertible it follows}\nonumber\\
    &=\prod_j
        \mathcal{D}^\dagger(\mathcal{J}_j)
        \underbrace{\mathrm{e}^{-i\tau_2\omega a_j^\dagger a_j}\mathcal{D}(\mathcal{J}_j)\mathrm{e}^{i \tau_2\omega a_j^\dagger a_j}}_\text{Baker-Campbell-Hausdorff}
        \cdot\exp\left\{-i\tau_2\omega \left(\eta^2\mathcal{P}_j-\mathcal{J}_j^\dagger\mathcal{J}_j\right)\right\}\\
    &=\prod_j
        \mathcal{D}^\dagger(\mathcal{J}_j)
        \exp{\mathcal{J}_j e^{-i\tau_2\omega}a_j^\dagger -\mathcal{J}_j^\dagger e^{i\tau_2\omega} a_j)}
        \cdot\exp\left\{-i\tau_2\omega \left(\eta^2\mathcal{P}_j-\mathcal{J}_j^\dagger\mathcal{J}_j\right)\right\}\\
    &=\prod_j
        \mathcal{D}(-\mathcal{J}_j)
        \mathcal{D}\left(\mathcal{J}_j e^{-i\tau_2\omega}\right)
        \cdot\exp\left\{-i\tau_2\omega \left(\eta^2\mathcal{P}_j-\mathcal{J}_j^\dagger\mathcal{J}_j\right)\right\}\\
    &=\prod_j\exp{-i\mathcal{J}_j^\dagger\mathcal{J}_j\sin(\omega\tau_2)-i\tau_2\omega \left(\eta^2\mathcal{P}_j-\mathcal{J}_j^\dagger\mathcal{J}_j\right)}
        \cdot\mathcal{D}\left(\mathcal{J}_j(e^{-i\omega\tau_2}-1)\right)\\
    &=\prod_j\exp{-i\tau_2\omega\left[
        \mathcal{J}_j^\dagger\mathcal{J}_j \left(\mathrm{sinc}(\omega\tau_2)-1\right)+\eta^2\mathcal{P}_j\right]}
        \cdot\mathcal{D}\left(\mathcal{J}_j(e^{-i\omega\tau_2}-1)\right)\\
    &=\prod_j\exp{-i\tau_2\omega\left[
        \left(\frac{\mathcal{G}_j^2}{\omega^2}+\eta^2\mathcal{P}_j\right)
        \left(\mathrm{sinc}(\omega\tau_2)-1\right)+\eta^2\mathcal{P}_j\right]}
        \cdot\mathcal{D}\left[\left(\frac{\mathcal{G}_j}{\omega}+i\eta \mathcal{P}_j\right)
        (e^{-i\omega\tau_2}-1)\right].
\end{align}
\endgroup
If we insert this result now in Eq.~(\ref{eq:time-evolution-2}), we obtain
\begin{align}
    \mathcal{U}_{\tau_1+\tau_2+2s}
    =\prod_j&
        \left(\dyad{\downarrow}_k-\mathcal{P}_j\right)
        \exp{-i\tau_2\omega\left[
            \left(\frac{\mathcal{G}_j^2}{\omega^2}+\eta^2\mathcal{P}_j\right)
            \left(\mathrm{sinc}(\omega\tau_2)-1\right)+\eta^2\mathcal{P}_j\right]}\nonumber\\
        &\cdot\exp{-i(\tau_1+\tau_2) \frac{\tau_2 V}{\tau_1+\tau_2} \mathcal{P}_j \mathcal{P}_{j+1}}\cdot\mathcal{D}\left[\left(\frac{\mathcal{G}_j}{\omega}+i\eta \mathcal{P}_j\right)
            (e^{-i\omega\tau_2}-1)\right]
        \cdot\exp{-i(\tau_1+\tau_2)\omega a_j^\dagger a_j}\\
        \nonumber\\
    =\prod_j&
            \left(\dyad{\downarrow}_k-\mathcal{P}_j\right)
            \exp{-i(\tau_1+\tau_2)\left[
                \frac{\tau_2 V}{\tau_1+\tau_2} \mathcal{P}_j \mathcal{P}_{j+1}
                +\frac{\mathcal{G}_j\tau_2}{\tau_1+\tau_2}\frac{\mathcal{G}_j}{\omega}\left(\mathrm{sinc}(\omega\tau_2)-1\right)
                +\frac{\sin{(\omega\tau_2)}}{\tau_1+\tau_2}\eta^2\mathcal{P}_j
            \right]}\nonumber\\
        &\cdot\mathcal{D}\left[\left(\frac{\mathcal{G}_j}{\omega}+i\eta \mathcal{P}_j\right)(e^{-i\omega\tau_2}-1)\right]
        \cdot
        \exp{-i(\tau_1+\tau_2)\omega a_j^\dagger a_j}\\
    =\prod_j&
            e^{-i\pi\mathcal P_j}
            \exp{-i(\tau_1+\tau_2)\left[
                \frac{\tau_2 V}{\tau_1+\tau_2} \mathcal{P}_j \mathcal{P}_{j+1}
                +\frac{\mathcal{G}_j\tau_2}{\tau_1+\tau_2}\frac{\mathcal{G}_j}{\omega}\left(\mathrm{sinc}(\omega\tau_2)-1\right)
                +\frac{\sin{(\omega\tau_2)}}{\tau_1+\tau_2}\eta^2\mathcal{P}_j
            \right]}\nonumber\\
        &\cdot\mathcal{D}\left[\left(\frac{\mathcal{G}_j}{\omega}+i\eta \mathcal{P}_j\right)(e^{-i\omega\tau_2}-1)\right]
        \cdot
        \exp{-i(\tau_1+\tau_2)\omega a_j^\dagger a_j}\\
    =\prod_j&
            \exp{-i(\tau_1+\tau_2)\left[
                \frac{\tau_2 V}{\tau_1+\tau_2} \mathcal{P}_j \mathcal{P}_{j+1}
                +\frac{\mathcal{G}_j\tau_2}{\tau_1+\tau_2}\frac{\mathcal{G}_j}{\omega}\left(\mathrm{sinc}(\omega\tau_2)-1\right)
                +\frac{\sin{(\omega\tau_2)}}{\tau_1+\tau_2}\eta^2\mathcal{P}_j
                +\frac{\pi}{\tau_1+\tau_2}\mathcal P_j
            \right]}\nonumber\\
        &\cdot\mathcal{D}\left[\left(\frac{\mathcal{G}_j}{\omega}+i\eta \mathcal{P}_j\right)(e^{-i\omega\tau_2}-1)\right]
        \cdot
        \exp{-i(\tau_1+\tau_2)\omega a_j^\dagger a_j}.
\end{align}
We now again utilize the short laser pulses introduced earlier, see Eq.~(\ref{eq:pi-pulse-approx}).
This allows us to write $T=\tau_1+\tau_2+2s\approx \tau_1+\tau_2$ to obtain the time evolution operator Eq.~(5) of the main text, 
\begin{align}
    \mathcal U(T)=\prod_j&\textcolor{AccentsRed4}{\underbrace{ \exp{-iT\left[
                \frac{\tau_2 V}{T} \mathcal{P}_j \mathcal{P}_{j+1}
                +\frac{\mathcal{G}_j\tau_2}{T}\frac{\mathcal{G}_j}{\omega}\left(\mathrm{sinc}(\omega\tau_2)-1\right)
                +\frac{\sin{(\omega\tau_2)}\eta^2+\pi}{T}\mathcal{P}_j
            \right]}}_\text{effective spin dynamics}}\nonumber\\
        &\cdot\textcolor{AccentsBlue5}{\underbrace{\mathcal{D}\left[\left(\frac{\mathcal{G}_j}{\omega}+i\eta \mathcal{P}_j\right)(e^{-i\omega\tau_2}-1)\right]}_\text{spin-motion coupling}}
        \cdot
        \textcolor{AccentsYellow4}{\underbrace{\exp{-iT\omega a_j^\dagger a_j}}_\text{free oscillator evolution}}.
\end{align}

\FloatBarrier
\subsection{Spin decoherence due to spin-motion coupling under ultra-strong laser pulses}
In this section, we investigate spin-motion coupling effects under ultra-strong laser pulses which are experimentally investigated in Ref. \cite{Bharti2024}.
Here, the laser pulses are characterized by a large Rabi frequency $\Omega_0\gg V$ that effectively overcomes the Rydberg interaction $V$.
This interaction is much stronger than the trap frequency with $V\gg\omega$.
As well as in the main text, each laser pulse is designed as a $\pi$~pulse with $\Omega_0 s=\pi/2$, resulting in a tiny pulse duration $s$.
Thus, for these ultra-strong pulses, the product $\omega\tau_2$ of the atomic oscillations becomes tiny, $\omega\tau_2\ll 1$. 

\begin{figure}[h]
    \centering
    \includegraphics{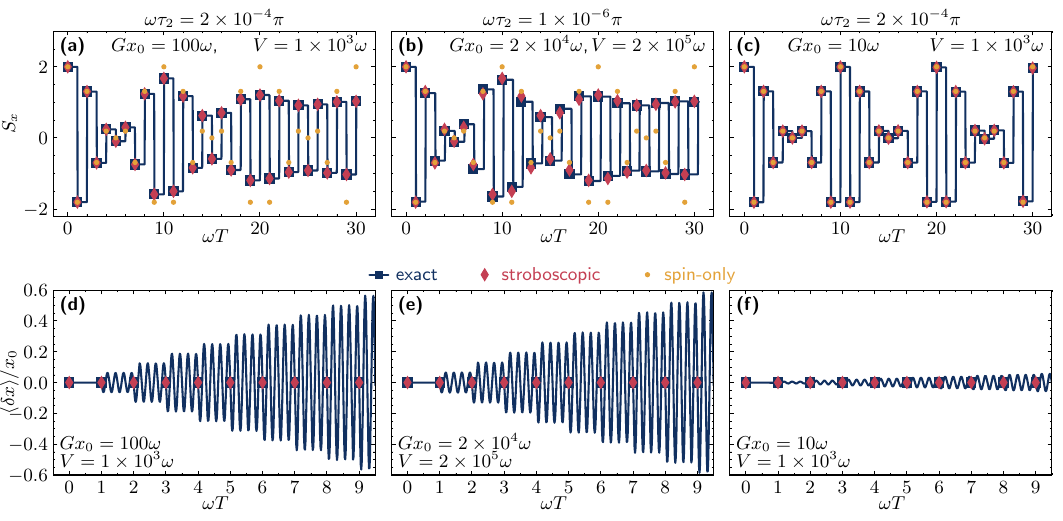}
    \caption{\textbf{Spin-motion coupling for ultra-strong laser pulses.}
    \textbf{(a)}~Exact- (blue) and effective spin dynamics (yellow) of $S_x$ deviate significantly.
    Under the effective spin dynamics (Eq.~(6) of the main text), periodic oscillations occur. However, due to spin-motion coupling, the exact dynamics (Eq.~(4) of the main text) show damped oscillations which we associate with spin decoherence. Here, we set $\omega\tau_2=2\times10^{-4}$ and the Rydberg interaction $V=1000\omega$.
    \textbf{(b)}~Scaling down $\tau_2$ and proportionally adjusting $V$ and $G$ cannot reduce spin decoherence
    in $S_x$ compared to panel (a).
    \textbf{(c)}~Reducing the effective gradient weakens spin decoherence, compared to (a) and (b). Thus, effective spin dynamics coincide with the exact dynamics.
    \textbf{(d)-(f)}~ The amplitude of the interatomic distance $\ev{\delta x}$ is mainly affected by the effective gradient $Gx_0\tau_2$ and impacts the observed spin decoherence.
    In all simulations the harmonic oscillator basis is truncated to contain a maximum of $5$ excitations, the parameters values are $T=8\pi/\omega$, $\eta = 0.6$ and $s = \pi/(2\cross 10^7\omega)$.
    }
    \label{fig:slight-out-of-phase}
\end{figure}

As shown in the main text, the system's time evolution operator $\mathcal U(T)$ simplifies (see Eq.~(5) in the main text) when strictly $\omega\tau_2=0$, indicating that spin and oscillatory motion decouple.
However, here we do not match this condition perfectly.
In order to understand the impact of residual spin-motion coupling in the limit $\omega\tau_2\ll1$, we perform the same study as in Fig.~2 of the main text.
Specifically, we show the time evolution of $S_x$ for small values of $\omega\tau_2$ in Fig.~\ref{fig:slight-out-of-phase}(a).
We compare the exact evolution of $S_x$ which includes spin-motion coupling (see Eq.~(4) of main text) with the effective spin dynamics calculated by Eq.~(6) of the main text.
We observe that deviations during time evolution occur. In particular, the effective spin dynamics lead to periodic oscillations, while the exact dynamics show damped oscillations. This mismatch of exact- and effective spin dynamics, which here results in damped oscillations, can be interpreted as spin decoherence, which arises due to the entanglement between oscillatory degrees of freedom and spins.
In Fig.~\ref{fig:slight-out-of-phase}(b), we reduce $\omega\tau_2$ by a factor of $200$ compared to panel (a), while proportionally increasing $V$ and $G$ to maintain the effective interaction strength $V_\mathrm{eff}=\tau_2V/T$ and the effective gradient operator $\mathcal G_{j,\mathrm{eff}}=\tau_2\mathcal G_j/T$.
These effective quantities occur in the effective Hamiltonian, see Eq.~(6) of the main text, and define the oscillation envelopes of $S_x$.
Compared to panel (a), we observe that the spin decoherence in $S_x$ does not decrease and deviations between exact- and effective spin dynamics are still present in panel (b).
This can be explained by expanding the collective displacement operator $\mathcal{J}_j$ (see main text). It couples spins and oscillatory motion and is the origin of the observed spin decoherence.
For $\omega\tau_2\ll 1$,
\begin{equation}
    \mathcal{J}_j\approx-i\tau_2\mathcal{G}_j+\omega\tau_2\eta \mathcal{P}_j=-i\mathcal{G}_{j,\mathrm{eff}}+\omega\tau_2\eta \mathcal{P}_j.
\end{equation}
As we show, $\mathcal J_j$ is dependent on the effective gradient operator $\mathcal G_{j,\mathrm{eff}}$.
Therefore, $\mathcal J_j$ remains unchanged for same $\mathcal G_{j,\mathrm{eff}}=\tau_2\mathcal{G}_{j}$.
This is why reducing $\omega\tau_2$ while increasing $G$ does not reduce spin decoherence.
On the other hand, reducing $\mathcal{G}_{j,\mathrm{eff}}$ without changing $\omega\tau_2$ reduces spin decoherence.
This means that by reducing the effective gradient, spin-motion coupling decreases, which ultimately removes spin decoherence as shown in Fig.~\ref{fig:slight-out-of-phase}(c).

\subsection{Decoupling spin-motion with a spin-motion echo}
In the main text, we examined how adjusting the oscillation phase, $\omega\tau_2$, allows to decouple spin-motion dynamics. Here, we explore a complementary approach using a spin-motion echo. This method is based on setting the total sequence duration to satisfy $T=n\pi/\omega$ with $n$ as an odd integer.
The spin-motion echo relies on the application of two successive dressing sequences that enable the decoupling.
Each sequence is describes by the time evolution operator $\mathcal U(T)$.
To illustrate, we evaluate the time evolution operator, see Eq.~(5) of the main text, for both successive sequences
\begin{align}
    \mathcal U^2(T=\frac{n\pi}{\omega})&=\prod_j    e^{-iT H_{j,\mathrm{eff}}}    \mathcal{D}(\mathcal{J}_j)    e^{-iT \omega a_j^\dagger a_j}
    \cdot e^{-iT H_{j,\mathrm{eff}}}    \mathcal{D}(\mathcal{J}_j)    e^{-iT \omega a_j^\dagger a_j}\nonumber\\
    &=\prod_j    e^{-2iT H_{j,\mathrm{eff}}}    \mathcal{D}(\mathcal{J}_j)    \underbrace{e^{-iT \omega a_j^\dagger a_j} \mathcal{D}(\mathcal{J}_j) e^{iT \omega a_j^\dagger a_j}}_\text{Baker-Campbell-Hausdorff}e^{-2iT \omega a_j^\dagger a_j}\nonumber\\
    &=\prod_j    e^{-2iT H_{j,\mathrm{eff}}}    \mathcal{D}(\mathcal{J}_j)\mathcal D(\mathcal J_j e^{-i\pi n})e^{-2iT \omega a_j^\dagger a_j}\nonumber\\
    &=\prod_j    e^{-2iT H_{j,\mathrm{eff}}}    \mathcal{D}(\underbrace{[1+ e^{-i\pi n}]}_{=0,\text{ if $n$ is odd}}\mathcal{J}_j)e^{-2iT \omega a_j^\dagger a_j}\nonumber\\
    &=\prod_j    e^{-2iT H_{j,\mathrm{eff}}} e^{-2iT \omega a_j^\dagger a_j}.
\end{align}
We see directly that spin-motion dynamics decouple under the chosen sequence duration for two successive applications of $\mathcal U(T)$.
This justifies the spin-motion decoupling for the spin-motion echo.

\FloatBarrier
\section{Analysis of the adiabatic dressing protocol}
\subsection{Adiabaticity and finite Rydberg lifetime: A case study}
During the adiabatic dressing sequence, we require the state to be smoothly connected to the instantaneous eigenstates. This preserves the population of all basis states $\mathcal C$ after an adiabatic sequence.
To quantify the adiabaticity of a sequence, we define the in-adiabaticity
\begin{equation}
    \mathcal{E}(T)=1-\max_\mathcal C \abs{\bra{\mathcal C}\mathcal{U}(T)\ket{\mathcal C}},
\end{equation}
where $\mathcal C$ is a basis state (configuration) of the Pauli $z$-basis.
We evaluate $\mathcal E$ for all basis states and take the maximum.
Importantly, $\mathcal E$ measures how adiabatic the dressing sequence is. For a completely adiabatic sequence, $\mathcal E=0$. As we show later, increasing the sequence duration $T$ improves the in-adiabaticity.

On the other hand, we excite Rydberg states coherently during the dressing sequence.
With increasing sequence duration $T$, the spontaneous decay probability $p$ increases as well, leading to higher decoherence.
The decay probability $p$ is approximately given by the time-integrated Rydberg excitation $\int_0^T\mathrm{d}t \sum_j \ev{n_j}(t)$ during the sequence with decay rate $\Gamma$ of Rydberg state \ry as
\begin{equation}
p(\Omega_0 T)=1-\exp\left(-\Gamma \int_0^T\mathrm{d}t \sum_j \ev{n_j}(t)\right)\approx \Gamma \int_0^T\mathrm{d}t \sum_j \ev{n_j}(t).
\end{equation}
In order for the system to stay coherent during a dressing sequence, $p\ll 1$ is required.
To demonstrate the trade-off between adiabaticity and decoherence due to sponatenoues decay, we study the system discussed in Fig.~3 of the main text in different parameter regimes, see Tab.~\ref{tab:case-study}.
\begin{table}[h]
    \begin{tabularx}{\textwidth}{l|XXXXX|XXX|X|X}
    \toprule
     & $\Omega_0/\Gamma$ & $\Delta_0/\Gamma$ & $\Delta_1/\Gamma$ & $V/\Gamma$ & $\Gamma T$ & $\varphi_\uparrow/(2\pi)$ & $\varphi_{\uparrow\uparrow}/(2\pi)$ & $\varphi_{\uparrow\uparrow\uparrow}/(2\pi)$ & $p$ & $\mathcal E$ \\
    \midrule
    (\textbf{1}) & 100 & 800 & 0 & 400 & 2.00 & 8.09 & 13.91 & 18.98 & $4.0\cross10^{-1}$    & $7.9\cross 10^{-08}$\\
    (\textbf{2}) & 100 & 800 & 200 & 400 & 2.00 & 3.93 & 7.58 & 11.04 & $1.8\cross10^{-1}$   & $1.6\cross 10^{-07}$\\
    (\textbf{3}) & 1000 & 8000 & 0 & 4000 & 0.20 & 8.09 & 13.91 & 18.98 & $4.0\cross10^{-2}$ & $7.9\cross 10^{-08}$\\
    (\textbf{4}) & 1000 & 8000 & 0 & 8000 & 0.20 & 8.09 & 13.41 & 17.94 & $3.5\cross10^{-2}$ & $2.3\cross 10^{-07}$\\
    (\textbf{5}) & 1000 & 8000 & 0 & 4000 & 0.05 & 4.05 & 6.96 & 9.49 & $9.9\cross10^{-3}$   & $1.6\cross 10^{-07}$\\
    (\textbf{6}) & 1000 & 8000 & 0 & 4000 & 0.01 & 0.45 & 0.73 & 0.98 & $2.2\cross10^{-3}$   & $4.3\cross 10^{-02}$\\
    \bottomrule
\end{tabularx}
\caption{\textbf{Adiabaticity vs. decoherence from spontaneous decay of Rydberg states.}
The rows ($\bm 1-\bm 6$) compare dressing sequences with different parameter sets with Rabi frequency $\Omega_0/\Gamma$, detunings $\Delta_{0,1}$, interaction strength $V$ and sequence duration $\Gamma T$ scaled by the decay rate $\Gamma$. For all sets, we calculate the dynamical phases $\varphi_\alpha$, see Eqs.~(13) of the main text, the in-adiabaticity $\mathcal E$ and the decay probability $p$.
($\bm 1$)~Resonant dressing sequence ($\Delta_1$=0) for same parameters as in Fig.~3(b) of main text. Here, Rydberg states can be excited resonantly.
($\bm 2$)~Compared to (1), for a dressing sequence which does not approach resonance ($\Delta_1=200\Gamma$), dynamical phases reduce.
($\bm 3$)~In comparison to case~(1), all parameters are scaled by increasing $\Omega_0$. This decreases the decay probability $p$.
($\bm 4$)~In comparison to case~(3), increasing the interaction strength $V$, leads to decreased dynamical phases but improved $\mathcal E$. This can be explained by the changed energy spectrum of the system.
($\bm 5$)~Compared to~(3), a decreased sequence duration $\Gamma T$ leads to lower $p$ but worse $\mathcal E$, where the dynamic phases decrease.
($\bm 6$)~A significantly longer sequence duration compared to~(5), lowers $p$ but increases the in-adiabaticity drastically.
}
\label{tab:case-study}
\end{table}

\FloatBarrier
\subsection{Relative strength of multi-body interactions}
\begin{figure}[h]
    \centering
    \includegraphics{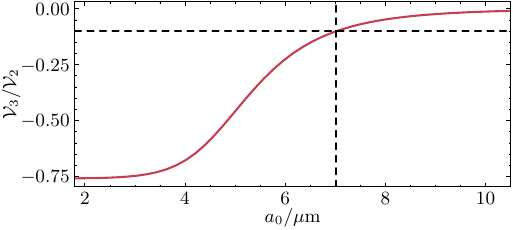}
    \caption{\textbf{Ratio of three-body and two-body interaction strength.}
    Ratio of the three-body and two-body interaction strength $\mathcal V_3/\mathcal V_2$ for the parameters shown in Fig.~3(d) of the main text.
    With decreasing distance $a_0$, the three-body interaction dominates the two-body one.
    Specifically, for distances $a_0\geq \SI{7}{\micro\meter}$, it is $\mathcal V_3\leq 10/\mathcal V_2$.
    }
    \label{fig:V2-vs-V3}
\end{figure}

The effective Hamiltonian of the adiabatic dressing sequence can be decomposed into contributions of $k$-body interactions.
Their relative strengths are affected by the distance dependent Rydberg interaction $V(a_0)$.
The three-body interaction strength $\mathcal V_3$ increases for shorter interatomic distance $a_0$.
Increasing $a_0$, the two-body interaction strength starts dominating. This is seen in Fig.~\ref{fig:V2-vs-V3}, where we show the ratio $\mathcal V_3/\mathcal V_2$.




\end{document}